\documentclass[letter]{ptptex}
\usepackage{wrapft}

\usepackage{graphicx}

\newcommand {\beq}{\begin{equation}}
\newcommand {\eeq}{\end{equation}}
\newcommand {\bea}{\begin{eqnarray}}
\newcommand {\eea}{\end{eqnarray}}
\newcommand {\nn}{\nonumber \\}
\newcommand {\Tr}{{\rm Tr\,}}

\newcommand {\m}{\mu}

\newcommand {\pl}{\partial}

\newcommand {\be}{\beta}

\newcommand {\La}{\Lambda}

\newcommand {\ep}{\epsilon}

\newcommand {\e} {\mbox{e}}

\newcommand {\del}  {\delta}

\newcommand {\half}{ {\frac{1}{2}} }

\def\overleftarrow#1{\vbox{\ialign{##\crcr
 $\leftarrow$\crcr\noalign{\kern-1pt\nointerlineskip}
 $\hfil\displaystyle{#1}\hfil$\crcr}}}

\newcommand {\ktil} {{\tilde k}}

\newcommand {\ptil}{{\tilde p}}

\newcommand {\intp} {{\int \frac{d^4p}{(2\pi)^4}}}

\newcommand {\ra} {\rightarrow}

\newcommand {\pr}   {{\quad .}}
\newcommand {\com}  {{\quad ,}}
\newcommand {\q}    {\quad}

\pubinfo{Vol.~, No.~,  2008}

\title{
Casimir and Vacuum Energy of 5D Warped System and 
Sphere Lattice Regularization
}

\author{
Shoichi \textsc{Ichinose}\footnote{
E-mail: ichinose@u-shizuoka-ken.ac.jp} 
}

\inst{
Laboratory of Physics, School of Food and Nutritional Sciences, 
University of Shizuoka\\
Yada 52-1, Shizuoka 422-8526, Japan
}

\recdate{
December 25, 2007}

\abst{
We examine the Casimir energy of 5D electro-magnetism 
in the recent standpoint. Z$_2$ symmetry is taken into account. 
After confirming the consistency with the past result, 
we do new things based on a new regularization. 
The regularization is based on the minimal area principle 
and the regularized configuration is  
the {\it sphere lattice}.  
We do it not in the Kaluza-Klein expanded form but in the
closed form. 
The formalism is based on the heat-kernel
approach using the position/momentum propagator. 
A useful expression of the Casimir energy, in terms of the P/M propagator, 
is obtained. 
Renormalization flow is realized as the change along 
the extra-axis. 
}

\begin{document}

\maketitle

The present common image about the compactification of the 
higher-dimensional model is strongly based on the work 
by Appelquist and Chodos\cite{AC83}. They considered 
the Kaluza-Klein model, which is the 5D unified model 
of the graviton, the photon and the dilaton. The starting Lagrangian 
is the pure 5D Einstein gravity on $S^1\times{\cal M}_4$. 
They took the standard approach of the quantum field theory
, the background field method, and calculated the Casimir energy
(taking the flat vacuum). 
After the appropriate regularization for the KK-expansion series expression, 
they obtain (for one scalar mode with the even parity)
\bea
V(l)=
\frac{1}{5}l\La^5
-\frac{3}{4}\frac{\zeta(5)}{l^4}\com\q
F(l)=-\frac{\pl V}{\pl l}=
-\frac{1}{5}\La^5
-3\frac{\zeta(5)}{l^5}
\com
\label{5dEM21}
\eea
where $l$ is the periodicity parameter ($y\ra y+2l$), and $\La$ is the 
4D momentum cut-off. 
The first term of $V(l)$ is {\it quintically} divergent. 
This quantity comes from the UV-divergences of 5D quantum fluctuation. 
Dropping the (divergent) constant term, the Casimir force is {\it finitely} 
obtained as $-3{\zeta(5)}/{l^5}$. 

In the closed form, $E_{Cas}$ of 5D electro-magnetism is expressed as
\bea
E_{Cas}(l)
=\intp\int_{p^2}^\infty\{\half\Tr G_k^-(y,y')+2\Tr G_k^+(y,y') \}dk^2
\equiv \frac{2\pi^2}{(2\pi)^4}\int_0^\infty d\ptil \int_0^l dy\ptil^3 F(\ptil,y)
\pr
\label{HK18}
\eea
where the P/M propagators are 
$G_k^\mp(y,y')=\pm\{\cosh \ktil(|y+y'|-l)\mp\cosh \ktil(|y-y'|-l)\}/
4\ktil \sinh\ktil l\ ,
-l\leq y\leq l,\ -l\leq y'\leq l,\ 
\ktil\equiv\sqrt{k_\m k^\m}\ ,\ k_\m k^\m>0
(\mbox{space-like})$. 
$F(\ptil,y)$ is expressed by the Gauss's hypergeometric function. 
The integral region of the equation (\ref{HK18}) is displayed in Fig.1. 
In the figure, we introduce the UV and IR regularization cut-offs, 
$\m=1/l\leq\ptil\leq\La$, $\ep=1/\La\leq y\leq l$.
From a close numerical analysis of ($\ptil,y$)-integral (\ref{HK18}), 
we have confirmed
$E_{Cas}(\La,l)=({2\pi^2}/{(2\pi)^4})\left\{ -0.1247 l\La^5-1.773\times 10^{-16}({1}/{l^4})
-1.253\times 10^{-15}l^{-4}{\ln (l\La)}\right\}
$. 
This is the (almost) same result of the previous one (1). 
The $\La^5$-divergence shows the notorious problem
of the higher dimensional theories. 
In spite of all efforts of the past literature, 
we have not succeeded 
in defining the higher-dimensional theories.\newline 
\begin{wrapfigure}{r}{6.6cm}
\caption{
Space of (y,${\tilde p}$) for the integration. 
}
\includegraphics[height=5cm]{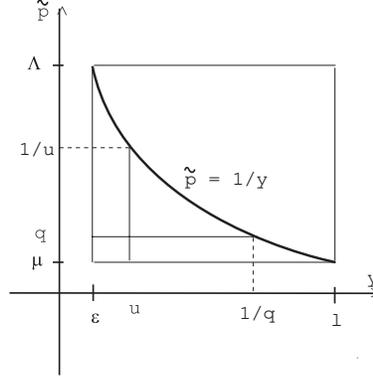}
\label{ypINTregion}
\end{wrapfigure}
The divergences 
cause problems. 
The famous example is 
the divergent cosmological constant in the gravity-involving theories.
\cite{AC83} 
We notice here we can avoid the divergence problem if we find a way to
{\it legitimately restrict the integral region in ($\ptil,y$)-space}. One 
proposal of this was presented by Randall and Schwartz\cite{RS01}. They introduced
the position-dependent cut-off,\ $\mu <\ptil <1/u\ ,\ u\in [\ep,l]$\ , 
for the 4D-momentum integral in the "brane" located at $y=u$. (See Fig.1)
The total integral region is the lower part of the hyperbolic curve $\ptil=1/y$. 
(They succeeded in obtaining the finite $\be$-function in the 5D warped vector
model.)  
Although they claim the holography is behind the procedure, 
the legitimateness of the restriction looks less obvious. We propose
an alternate one \cite{IM0703}
and give a legitimate explanation within the 5D QFT. 

\begin{wrapfigure}{r}{6.6cm}
\caption{
Space of (y,$\ptil$) for the integration (present proposal). 
}
\includegraphics[height=5cm]{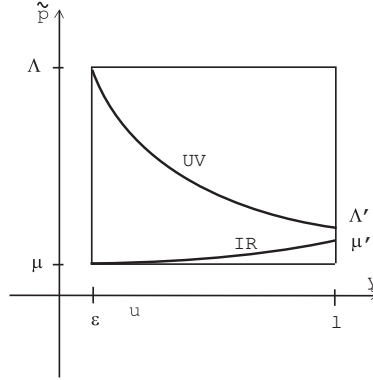}
\label{ypINTregion2}
\end{wrapfigure}
On the "3-brane" at $y=\ep$, we introduce the IR-cutoff $\mu$ and 
the UV-cutoff $\La$\ ($\mu\ll\La$). See Fig.2.  This is legitimate in the sense that we
usually do this procedure in the 4D {\it renormalizable} theories. On the
"3-brane" at $y=l$, we have another set of IR and UV-cutoffs, 
$\mu'$ and $\La'$. 
We consider 
the case: $\mu'\leq\La',\ \mu\sim\mu', \La'\ll\La$. This case leads to allow us to
introduce the renormalization flow. (See the later explanation of Fig.3.)
We claim here,  
as for the "3-brane" located at each point $y$ ($\ep<y<l$), the regularization 
parameters are determined by the {\it minimal area principle}. 
To explain it, we depict the regularization configuration(Fig.2) in the 5D coordinate space ($x^\m,y$) 
in Fig.3. 
The 5D volume region bounded by $B_{UV}$ and $B_{IR}$ is the integral region 
of the Casimir energy $E_{Cas}$. 
The forms of $r_{UV}(y)$ and $r_{IR}(y)$ can be
determined by the {\it minimal area principle}.
\bea
\del (\mbox{Surface Area})=0\com\q 
3-\frac{r\frac{d^2r}{dy^2}}{1+(\frac{dr}{dy})^2}=0\com\q 0\leq y\leq l
\label{surf2}
\eea
We have confirmed that there exist the solutions (geodesic curves) 
with the properties shown in Fig.2 or Fig.3 when we take appropriate
boundary conditions: $r(y=\ep), r(y=l)$. 

Instead of restricting the integral region, we have another approach to 
suppress UV and IR divergences. We introduce a {\it weight function} $W(\ptil,y)$.

\bea
E^W_{Cas}(l)\equiv\intp\int_0^ldy~ W(\ptil,y)F(\ptil,y)\com
\label{uncert1}
\eea
As the examples of $W$, we present $\e^{-l^2\ptil^2/2-(y^2/2l^2)}\equiv W_1(\ptil,y)$({elliptic suppression})
and $\e^{-\ptil y}\equiv W_2(\ptil,y)$ ({hyperbolic suppression}). 
We have evaluated the divergence behaviour of $E^W_{Cas}$ by 
numerically performing the $(\ptil,y)$-integral (\ref{uncert1}) for 
the rectangle region of Fig.1.  
\bea
E^W_{Cas}=\left\{
\begin{array}{cc}
\frac{c_{10}}{l^4}-21.4\frac{\La}{l^3}+c_{11}\frac{\La\ln\La}{l^3} & \mbox{for}\q W_1(\ptil,y) \\
-\frac{c_{20}}{l^4}-0.216\La^4+c_{21}\La^4\ln\La  &\mbox{for}\q W_2(\ptil,y)
\end{array}
           \right.
\label{uncert1b}
\eea
where $c's$ are unstable for the range $l=(10,20,40),\La=10\sim 10^3$ and are given by 
$c_{10}=(26.3,18.2,10.0)$, $c_{11}=(5.52,2.78,1.39)\times 10^{-3}$, 
$c_{20}=(6.47,55.6,446)\times 10^8$, $c_{21}=(4.73,2.35,1.18)\times 10^{-5}$. In particular 
$c_{11}$ and $c_{21}$ changes as $1/l$ (decreases as $l$ increases).
\begin{wrapfigure}{r}{6.6cm}
\caption{
Regularization Surface $B_{IR}$ and $B_{UV}$ in the 5D coordinate space $(x^\m,y)$,
 Flow of Coarse Graining (Renormalization) and Sphere Lattice. 
}
\includegraphics[height=8cm]{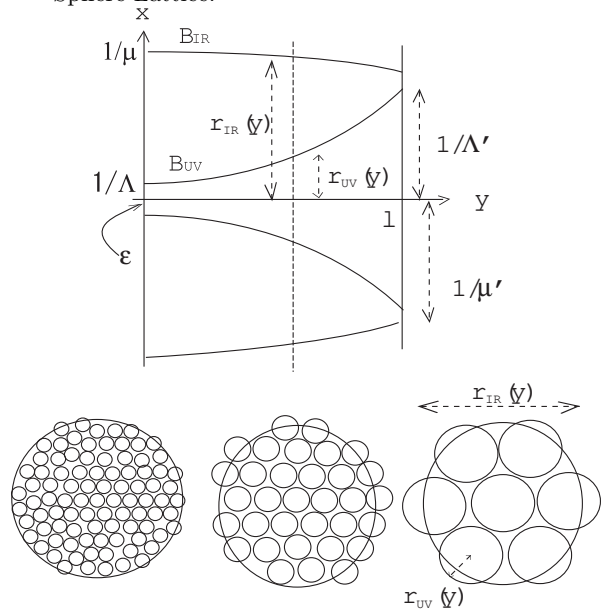}
\label{IRUVRegSurf}
\end{wrapfigure}
$W_2$ corresponds to the restriction approach by Randall-Schwartz and the above 
result is consistent with theirs. Its suppression, however, is insufficient. 
$W_1$ gives, after normalization by the factor $l\La$, the desired 
log-divergence. In this case, the Casimir energy is finitely obtained by 
the {\it renormalization of the periodicity $l$}.  
\bea
-\frac{3}{4}\frac{\zeta(5)}{l^4}(1-4c \ln(l\La))=-\frac{3}{4}\frac{\zeta(5)}{{l}'^4},\nn
\frac{\pl}{\pl (\ln \La)}\ln\frac{l'}{l}=c~(\mbox{anom. dim.})
\pr
\label{conc1}
\eea
Fig.3 shows the renormalization flow. For interacting theories, such as 
5D YM theories, 
the scaling of the renormalized coupling $g(y)$ is given by
\bea
\be
=-\frac{1}{3}\frac{1}{\frac{\pl}{\pl y}\ln r(y)}\frac{1}{g}\frac{\pl g}{\pl y},
\label{uncert4}
\eea
where $g(y)$ is a renormalized coupling at $y$ and $r(y)$ is an appropriate geodesic.

Finally we comment on the meaning of the weight function. First 
we can define it by requiring that the dominant contribution to $E_{Cas}$ 
(\ref{uncert1}), which is obtained by the steepest-descend method to (\ref{uncert1}), 
coincides with the geodesic curve, which is obtained by the minimal area principle 
for the surface in the bulk. 
\bea
\frac{d\ptil}{d y}=
\frac{   -\frac{\pl\ln (WF)}{\pl y}   }
     {    \frac{3}{\ptil} +\frac{\pl\ln (WF)}{\pl \ptil} }
\pr
\label{uncert3}
\eea
Secondly, we notice the $W_1$ is the harmonic oscillator Hamiltonian, 
hence the weighted system can be regarded as some quatum mechanical 
system of 5D-space coordinates where the extra coordinate $y$ and the absolute value 
of the 4D momentum $\ptil$ are in the conjugate relation 
(new uncertainty principle). 

So far the flat geometry is considered. For the warped case, the similar analysis 
can be done. One additional parameter (curvature parameter of AdS$_5$), besides $l$, 
comes into the arguments. We are also examining the vacuum energy and the self energy 
using interacting theories such as 5D $\Phi^4$-theory and 5D YM.


\begin{thebibliography}{99}
\bibitem{AC83}   
T. Appelquist and A. Chodos, Phys.\ Rev.\ \textbf{D28} (1983) 772\\
T. Appelquist and A. Chodos, Phys.\ Rev.\ Lett.\ \textbf{50} (1983) 141
\bibitem{RS01}   
L. Randall and M.D. Schwartz, \JHEP{0111,2001,003}
, hep-th/0108114
\bibitem{IM0703}  
S. Ichinose and A. Murayama, Phys.\ Rev.\ \textbf{D76}(2007),065008, hep-th/0703228. 
\end{thebibliography}
\end{document}